\documentclass[aps,twocolumn,letterpaper, nofootinbib, floatfix, superscriptaddress,amsmath,amssymb]{revtex4-2}
\usepackage{amsmath,amssymb, graphicx}
\usepackage{bm,times, color}
\usepackage[utf8]{inputenc}
\usepackage[colorlinks,breaklinks]{hyperref}
\usepackage[svgnames]{xcolor}
\hypersetup{linkcolor=DarkBlue, citecolor=DarkBlue, filecolor=black, urlcolor=DarkBlue}
\usepackage{comment, mathrsfs,slashed}
\usepackage{amsfonts}
\usepackage{bbm}
\usepackage[normalem]{ulem}
\def \p{\partial}
\def \dag{\dagger}
\def \mb{\mathbf}

\def \lan{\langle}
\def \ran{\rangle}

\def \ga{\gamma}

\begin{document}
\title{Chiral anomaly in (1+1) dimensions revisited: complementary kinetic perspective and universality}

\author{Wei-Han Hsiao}
\noaffiliation{}

\author{Chiao-Hsuan Wang}
\affiliation{Department of Physics and Center for Theoretical Physics, National Taiwan University, Taipei, 10617, Taiwan}
\affiliation{Center for Quantum Science and Engineering, National Taiwan University, Taipei, 10617, Taiwan}
\affiliation{Physics Division, National Center for Theoretical Sciences, Taipei, 10617, Taiwan}
\affiliation{Pritzker School of Molecular Engineering, University of Chicago, Chicago, IL, 60637, USA}
\date{November 2024}

\begin{abstract}
We reinvestigate the classic example of the chiral anomaly in (1+1) dimensional spacetime. By reviewing the derivation of charge conservation using the semiclassical Boltzmann equation, we show that chiral anomalies could emerge in (1+1) dimensions without Berry curvature corrections to the kinetic theory. The pivotal step depends only on the asymptotic behavior of the distribution function of the quasiparticle--and thus its dispersion relation--in the limit of $\mathbf p \to \pm\infty$ rather than the detailed functional form of the dispersion. We address two subjects motivated by this observation. First, we reformulate (1+1)-dimensional chiral anomaly using kinetic theory with the current algebra approach and the gradient expansion of the Dirac Lagrangian, adding a complementary perspective to existing approaches. Second, we demonstrate the universality of the chiral anomaly across various quasiparticle dispersions. For two-band models linear in the temporal derivative, with Fujikawa's method we show it is sufficient to have a chirality-odd strictly monotonic dispersion in order to exhibit the chiral anomaly.
\end{abstract}

\maketitle

\section{Introduction}
Anomaly refers to the breakdown of a symmetry in a classical model due to quantization. One of the most celebrated and vintage examples is the chiral anomaly of the (3+1) dimensional massless Dirac fermion in an electromagnetic field \cite{PhysRev.177.2426, Bell1969}. Classically, the two species of Weyl fermions in a massless Dirac fermion decouple. The chiral current $J_5^{\mu}$, defined as the difference between the currents of the Weyl fermions, is thus conserved $(\p_{\mu}J_5^{\mu} = 0)$. Nonetheless, quantum mechanically, $J^{\mu}_5$ cannot be conserved if the total U(1) current $J^{\mu}$, the sum of the two Weyl currents, is computed in a conserved fashion. Historically, this anomaly explains the finite decay rate of a pion into photons $(\pi^0 \to \ga\ga)$, and is dubbed the triangle anomaly because it appears in a triangle loop diagram when computing this physical process \cite{Bell1969}. Since its discovery in perturbative calculations, enormous efforts have been dedicated to understanding it, providing insights from both physical and mathematical perspectives. Examples include Fujikawa's path integral measure \cite{PhysRevLett.42.1195, PhysRevD.21.2848, Fujikawa2004} and the index theorem \`a la Atiyah and Singer \cite{10.2307/1970715, Atiyah1973}. The kinetic equation approach for the triangle anomaly emerged rather late, in the last decade \cite{PhysRevLett.109.162001, PhysRevLett.109.181602, PhysRevD.87.085016, PhysRevLett.110.262301}. In this formulation, the Berry curvature of a Weyl fermion around its Fermi surface induces deformations in the phase space volume measure and modifies the equal-time commutation relation between two density operators, leading to the chiral anomaly. This method provides a concise and elegant derivation for the negative magnetoresistance of the Weyl semimetals \cite{PhysRevB.88.104412, RevModPhys.90.015001}.

This work concentrates on a less popular but technically much simpler model that exhibits chiral anomaly: the (1+1) dimensional Dirac field in the presence of background electromagnetism. We will avoid calling it {\it the Schwinger model} \cite{PhysRev.128.2425}, as we do not plan on introducing dynamical electromagnetism. It is often considered a warm-up exercise of the triangle anomaly in most educational literature \cite{Bertlmann2000, Fujikawa2004, Peskin:257493}. Nevertheless, there are subtle differences that prevent us from applying every result from the studies on triangle anomaly to this lower dimensional counterpart and vice versa. For instance, the divergence in the calculation of the former occurs in the ultraviolet regime, whereas the ambiguity of the latter pertains to the proper definition of the charge in the ground state, an infrared quantity. Furthermore, the chiral kinetic theory cannot be defined in parallel because of the nonexistence of Berry curvature due to the dimensional constraint. One of the theses of this work is that a kinetic approach should still exist to bridge the fully quantum electrodynamic treatment and the heuristic argument of spectral flow, where charges are pumped out of or dumped into the Dirac sea depending on their chirality \cite{Bertlmann2000, Fujikawa2004, Peskin:257493, MANTON1985220, 1999hep.th....3255J}. Consequently, this work complements the earlier insight in (3+1) dimensions \cite{PhysRevLett.109.181602, PhysRevD.87.085016, PhysRevLett.110.262301}. 

To see how chiral anomaly can arise from an ordinary kinetic theory without geometric phases, let us consider the general Boltzmann equation:
\begin{align}
\label{boltzmann}\frac{\p n_{\mb p}}{\p t} + \mb v\cdot\frac{\p n_{\mb p}}{\p \mb x} + \mb f \cdot\frac{\p n_{\mb p}}{\p\mb p} = I[\delta n_{\mb p}],
\end{align}
where $n_{\mb p}(t, \mb x)$, $\mb v$, $\mb f$ and $I$ stand for the standard quasiparticle distribution function in the phase space $(\mb x, \mb p)$, quasiparticle velocity, external force, and collision integral, respectively. The conservation laws can be derived by computing the moment averages of $\mb p$. For instance, the charge conservation $\p_t\rho + \nabla\cdot\mb j = 0$ is obtained by integrating the equation with the measure $(d\mb p) := \int \frac{d^d\mb p}{(2\pi)^d}$ together with the identifications $\rho = \int (d\mb p)n_{\mb p}$ and $\mb j = \int (d\mb p) \mb vn_{\mb p}$. The external force term would vanishes under the standard integration by part argument \cite{lifshitz1995physical}. Such consideration is valid for most models and band structures where the regions of large momenta are filled or empty simultaneously and for models whose quasiparticle energy depend only on the magnitude of momentum. It could, however, be violated by an anomalous model such as one chiral sector of the Dirac Lagrangian in (1+1) dimensions. Taking the right-moving fermion, $n_{\mb p}|^{\infty}_{-\infty} = -1$, we consequently have the broken conservation law:
\begin{align}
\p_t \rho_R+ \p_x j_R = \frac{f}{2\pi},
\end{align}
which becomes anomalous provided $f = E$, the external electric force. By the same token we can easily see $\p_t \rho_L + \p_x j_L = -E/(2\pi)$ and therefore the overall chiral current obeys the anomaly equation 
\begin{align}\label{anomaly_0}
\p_{\mu}j^{\mu}_5 = \frac{1}{2\pi}\epsilon_{\mu\nu}F^{\mu\nu}.
\end{align}
In light of this observation, two topics will be addressed in this work: 
\begin{enumerate}
\item We first reformulate (1+1) dimensional chiral anomaly in terms of kinetic theory. The first step involves demonstrating the Hamiltonian formulation of the kinetic theory introduced in Ref.\onlinecite{PhysRevLett.109.181602}, together with the asymptotic behavior argument presented above, naturally implying the central extension of Kac-Moody algebra for the charge densities of chiral fermions. This extension, in turn, implies chiral anomaly. We also consider the conventional gradient expansion of the field theoretical correlation functions, showing that each chiral sector of the (1+1) dimensional Dirac fermion satisfies a chirality-dependent semiclassical Boltzmann equation. These exercises not only offer an alternative perspective filling the gap between the illustrative spectral-flow picture and other fairly technical approaches such as direct diagrammatics, dispersion relations, or Fujikawa's determinant, but also serve as the analogue of the kinetic equation approach to the (3+1) dimensional triangle anomaly. This approach provides straightforward derivations for anomalous currents in (1+1) dimensions.
\item Since this distribution function argument depends only on the asymptotic behavior of $n_{\mb p}$ as $\mb p \to \pm \infty$, a strictly monotonic dispersion relation would suffice to yield a finite difference at $T = 0$. To quantify this statement, we will consider a class of (1+1) dimensional Dirac-like models. They are linear in time derivative but allow generic functional forms of the spatial derivative. We show quantum mechanically such models have chiral anomaly so long as its chiral sectors assume the dispersion relations 
\begin{align*}
\omega =  {\epsilon_{\mb p} \sim \mb p^{2n-1}}, n\in\mathbb N.
\end{align*}
The leading contribution is the same value as that given by the Dirac field, $E/(2\pi)$, regardless of the value of $n$. The argument is then generalized to apply to general strictly monotonic dispersions $\omega \sim \epsilon(-i\p_x)$.
\end{enumerate}

These themes will be elaborated further in Secs. \ref{kinetic_sec} and \ref{dispersion_section} respectively. A couple of comments will follow to conclude the work.

\section{Kinetic formulation for (1+1) dimensional chiral anomaly}\label{kinetic_sec}
\subsection{Hamiltonian formulation for anomalous charge commutation relation}
The kinetic theory of fermionic quasiparticles is reformulated using the Hamiltonian functional $H[n_{\mb p}(\mb x)]$ and the commutation relation between the distribution functions $[n_{\mb p}(\mb x), n_{\mb q}(\mb y)]$. To derive Eq~\eqref{boltzmann} via the Louville equation $\p_t n_{\mb p}(\mb x) = i[H, n_{\mb p}(\mb x)]$ in the $(d+1)$ dimensional spacetime, the commutation relation is postulated to be 
\begin{align}
& i[n_{\mb p}(\mb x), n_{\mb q}(\mb y)] \notag\\
\label{commutation}=& (2\pi)^d\frac{\p}{\p \mb p}\delta(\mb p-\mb q)\cdot\frac{\p}{\p \mb x}\delta(\mb x-\mb y) [n_{\mb p}(\mb y) - n_{\mb q}(\mb x)]
\end{align}
in Ref.\onlinecite{PhysRevLett.109.181602}. The bulk of that seminal work is devoted to showing the equal-time commutator between two charge-density operators $ {\rho(\mb x) = \int d\Gamma n_{\mb p}(\mb x)}$ is anomalous when accounting for the effect of Berry curvature on the volume measure $d\Gamma$. Consider a model consisting of two species of fermions. In the free-theory limit,  {quasiparticle dispersions are given by $\omega = \epsilon_{\mb p} =  \sigma |\mb p|$ for $d>1$ and $\omega = \epsilon_{p_x}= \sigma p_x$ for $d = 1$}, where $\sigma =\pm 1$ denotes the {\it chirality}. Through a straightforward evaluation, it can be shown that in $(1+1)$ dimensions, Eq.~\eqref{commutation} naturally leads to anomalous density-density commutation relation for a chiral fermion without Berry curvature correction to the measure $d\Gamma$:
\begin{align}
& {[\rho_{\sigma}(x), \rho_{\sigma'}(y)]} = \int \frac{dp_x}{(2\pi)}\frac{dq_x}{(2\pi)}[n_{\sigma p}(x), n_{\sigma'q}(y)] \notag\\
 = & \frac{i}{2\pi}\delta_{\sigma\sigma'}\frac{\p}{\p x}\delta(x-y)n_{\sigma p}(y) {\Big|^{p_x = \infty}_{p_x = -\infty}}.
\end{align}
The subtlety of $d = (1+1)$ highlighted by this work is that $n_{\sigma p}(y) {|^{p_x = \infty}_{p_x = -\infty}}$ can be non-vanishing. Accounting for the different asymptotic behaviors of $n_{\sigma p}$ as we did in the introduction, when the quasiparticle dispersion is odd in momentum,  {$n_{\sigma p}(y)|^{p_x = \infty}_{p_x = -\infty} \sim -\sigma$}, the above reduces to the extended Kac-Moody algebra:
\begin{align}
\label{Kac-Moody}  {[\rho_{\sigma}(x), \rho_{\sigma'}(y)]} = \frac{\sigma}{2\pi i} \delta_{\sigma\sigma'}\p_x\delta(x-y).
\end{align}
Given this commutation relation, the anomaly can be derived in the same fashion as the hydrodynamic approaches of chiral Luttinger liquids \cite{PhysRevB.55.15832}. We consider an inhomogeneous electric potential  {$V(x)$} in the Hamiltonian:
\begin{align}
H = H_0 + \int dx\, \rho(x) {V(x)}.
\end{align}
$H_0$ is assumed to be a free Hamiltonian:  {$\sum_{\sigma}\int \frac{dp_x}{2\pi}\epsilon_{p_x} n_{\sigma p_x}$ with $\epsilon_{p_x}$ being its quasiparticle energy.} Although this formulation independent of spacetime dimensions, physically, the well-defined quasiparticles are destroyed in (1+1) dimensions as the short-range interaction is turned on.
Evaluating the commutator  {$[H, \rho_{\sigma}(x)]$}, the $H_0$ part yields the current conservation. The additional contribution from the potential reads
\begin{align}
 {\p_t\rho_{\sigma}} + \p_x j_{\sigma} =& i\int dy\,  {V(y)[\rho_{\sigma}(y), \rho_{\sigma}(x)]} \notag\\
\label{result1}=& -\frac{\sigma}{2\pi}\p_x {V} = \frac{\sigma}{2\pi}E.
\end{align}
We then recover the anomalous conservation of charge for chiral fermions. Charge conservation and chiral anomaly can be derived by taking the sum and the difference of the $\sigma = \pm 1$ sectors as we described earlier. Note that this derivation is independent of monopole or Berry phase and therefore should not be sensitive to adiabatic assumptions, and the current algebra \eqref{Kac-Moody} is unsurprisingly identical to those given by bosonization and the Bjorken-Johnson-Low (BJL) prescription \cite{Fujikawa2004, PhysRevD.97.016018}. 
\subsection{kinetic equation from microscopic models}
We next derive this result by evaluating the semi-classical expansion of the Dirac fermion in (1+1) dimensions. 
\begin{align}
\label{model}\mathscr L = i\bar{\psi}\gamma^{\mu}D_{\mu}\psi + \mu\psi^{\dag}\psi
\end{align}
with $\ga^{\mu} = (\sigma^y, i\sigma^x)$ and $D_{\mu} = \p_{\mu} + iA_{\mu}$. The chiral operator $\gamma_5 $ is  $\gamma^0\gamma^1 = \sigma^z$. In the absence of the fermion mass, the chiral sectors, or the upper and the lower entries of the spinor $\psi$, decouple. 
As pointed out earlier, a significant amount of effort has been dedicated to the kinetic theory formulation of the triangle anomaly in (3+1) dimensions. Nevertheless, not all arguments and techniques there can be adapted when tackling its (1+1) dimensional counterpart such as the Berry curvature term emerging in the hard-density loop expansions. As such, here we simply perform the most generic semi-classical approximation on each chiral sectors \cite{lifshitz1995physical, mah00, kamenev_2011}.
To proceed, we rephrase the Lagrangian~\eqref{model} as the sum of two chiral sectors:
\begin{align}
\mathscr L = \sum_{\sigma = \pm }\mathscr L_{\sigma},\ \mathscr L_{\sigma} = i\psi^{\dag}_{\sigma}[D_t + \sigma D_{x}]\psi_{\sigma} + \mu\psi^{\dag}_{\sigma}\psi_{\sigma}.
\end{align}  
The classical equations of motion are given trivially by varying $\psi^{\dag}$ and $\psi$, which are:
\begin{align}
i \p_t \psi_{\sigma} = -i \sigma\p_x\psi + V_{\sigma}\psi_{\sigma}, V_{\sigma} = A_0 + \sigma A_1 -\mu,
\end{align}
and its complex conjugate. 
To derive the kinetic equations, let us consider the Green's function 
\begin{align}
iG_{\sigma}(x_1, x_2) = \lan T\psi_{\sigma}(x_1)\psi^{\dag}_{\sigma}(x_2)\ran.
\end{align}
Here the time-ordered propagator at zero temperature is adopted, but the technique is general and should apply to any kind of Green's functions. Differentiating the Green's function with respect to $x$ and $x'$, and substituting the equations of motion for the time derivatives of the field operators, we obtain:
\begin{subequations}
\begin{align}
\label{left}(\p_{t_1} +  \sigma\p_{x_1} + i V_{\sigma}(x_1) ) iG_{\sigma}(x_1, x_2)  &= \delta(x_1-x_2),\\
\label{right} (-\p_{t_2} -\sigma\p_{x_2} +i V_{\sigma}(x_2)) iG_{\sigma}(x_1, x_2) &= \delta(x_1-x_2).
\end{align}
\end{subequations}
The sum and the difference of these two equations yield the on-shell condition and quantum kinetic equation, respectively. Next, we introduce the center of mass and relative coordinates $( {X^{\mu}}, s^{\mu})$:
\begin{subequations}
\begin{align}
&  {X} = \frac{x_1 + x_2}{2}\\
& s = x_1 - x_2.
\end{align}
\end{subequations}
In terms of these coordinates, the sum of Eqs.~\eqref{left} and~\eqref{right} reads:
\begin{align}
\left( i\frac{\p}{\p s^0} + i\frac{\p}{\p s^1}\sigma -V_{\sigma}( {X})\right)i G_{\sigma}= i\delta(s),
\end{align}
where we have expanded $V_{\sigma}(X\pm s/2)$ to the first order in $s$. Fourier transforming with respect to $s^{\mu}$ suggests $iG$ has a pole structure $i(\omega - \sigma p_x -V_{\sigma})^{-1}$. In the same expansion scheme, we consider the difference of Eqs.~\eqref{left} and~\eqref{right}, 
\begin{align}
\label{DiffLeftRight}\left(i \frac{\p}{\p  {X^0}} + i\sigma \frac{\p}{\p  {X^1}} - s^{\mu}\p_{\mu}V_{\sigma}\right) iG_{\sigma} = 0.
\end{align}
Schematically, this equation reduces to the kinetic equation after the Wigner transformation. To maintain the gauge covariance, let us consider the covariant Wigner-transformation:
\begin{align}
& i\mathscr G( {X}, p) = \int d^2s\, e^{isp}U( {X, X_+})iG( {X_+, X_-})U( {X_-, X}), \\
&  {s\cdot p = s^0\omega - s^1p_x,}\notag\\
& U(x, y) = \exp\left( i \int_{x}^y d\ell^{\mu}A_{\mu}\right),\  {X_{\pm} = X}\pm \frac{s}{2}.\notag
\end{align}
By plugging the inverse transformation of the above into Eq.~\eqref{DiffLeftRight}, we obtain the quantum Boltzmann equation:
\begin{align}
\left(\p_t  + \p_x \sigma + E\frac{\p}{\p p_x} + E\frac{\p}{\p\omega}\sigma \right)i\mathscr G_{\sigma}( {X}, p) =0
\end{align}
with the electric field $E = \p_tA_x - \p_x A_0$  {and $X^{\mu} = (t, x)$}. The third and fourth term have the obvious physical meanings of external force and power $\mb F\cdot\mb v$ respectively. 
The semi-classical Boltzmann equation follows by putting the quantum version on-shell via the on-shell condition $i\mathscr G_{\sigma}( {X}, p) = 2\pi\delta(\omega - \epsilon_{\sigma p}( {X}))n_{\sigma p}( {X})$ \cite{kamenev_2011}. We thus have the (1+1)-dimensional version of Eq.~\eqref{boltzmann}:
\begin{align}
\label{boltzmannKinetic}\frac{\p}{\p t} n_{\sigma p} + \frac{\p}{\p x}\sigma n_{\sigma p} + E\frac{\p}{\p  {p_x}}n_{\sigma p} = 0.
\end{align}
Let us now integrate over the last momentum variable with the measure $\int \frac{d { p_x}}{2\pi}$. The first two terms are simply $\p_t { \rho_{\sigma}} + \p_x j_{\sigma}$. To evaluate the third term, we note that at equilibrium, the sign of the quantity determines $\sigma  {p_x} + V_{\sigma}$ if the quasiparticle state is occupied. Away from equilibrium, this should still entail asymptotic occupation at $ {p_x}\to \pm \infty$, and accordingly:
\begin{align}
\int_{-\infty}^{\infty} \frac{d {p_x}}{2\pi}\frac{\p}{\p  {p_x}}n_{\sigma p} = -\frac{\sigma}{2\pi}.
\end{align}
Putting them altogether, we reproduce Eq.\eqref{result1}:
\begin{align}
\label{main_result}\p_t  {\rho_{\sigma}} + \p_x j_{\sigma} = \frac{\sigma}{2\pi}E.
\end{align}
 {
\subsection{anomalous number and energy currents} 
A natural question to explore is the application of the kinetic approach. In (3+1) dimensions, this approach offers elegant derivations for the chiral magnetic and vortical effects \cite{PhysRevLett.109.162001, PhysRevLett.109.181602}. We will conclude this section by demonstrating how, in (1+1) dimensions, the kinetic approach provides straightforward derivations for anomalous number and energy currents \cite{PhysRevLett.81.3503}.

Let us first consider the anomalous number current. From Eqs.~\eqref{boltzmann},~\eqref{boltzmannKinetic}, and~\eqref{main_result}, we can infer:
\begin{align}
j_{\sigma} = \int^{\infty}_{-\infty} \frac{dp_x}{2\pi}\sigma n_{\sigma p} - \bar j_{\sigma},
\end{align}
where $\bar j_{\sigma}$ represents the ground-state value, which is fixed by normal-ordering the ground state such that $j_{\sigma}(\mu_{\sigma}  = 0) = 0,$ where $\mu_{\sigma}$ is the constant chemical potential for chirality $\sigma$. At thermal equilibrium with temperature $T$, the distribution function is given by:
\begin{align}
n_{\sigma p} = \frac{1}{e^{(\sigma p_x - \mu_{\sigma})/T} + 1}.
\end{align}
We can straightforwardly compute:
\begin{align}
\frac{j_{\sigma}}{\sigma} = \int^{\infty}_{-\infty} \frac{dp_x}{2\pi} \left( \frac{1}{e^{(\sigma p_x-\mu_{\sigma})/T} + 1} - \frac{1}{e^{\sigma p_x/T} + 1}\right) = \frac{\mu_{\sigma}}{2\pi}.
\end{align}
Hence, the total current is:
\begin{align}
j = \sum_{\sigma} j_{\sigma} = \frac{1}{2\pi}(\mu_+ - \mu_-).
\end{align}
This can alternatively be derived by integrating Eq.~\eqref{main_result} with a static external electric field.

To investigate the anomalous energy current, we consider an equilibrium state with no external electromagnetic field $E = 0$ and $\mu_{\sigma} = 0$, allowing each species of fermions to have its own equilibrium temperature $T_{\sigma}$. We multiply Eq.~\eqref{boltzmannKinetic} by its quasiparticle energy $\epsilon_{p_x} = \sigma p_x$ and then integrate over $p_x$. These steps yield the conservation of energy density $\rho_{\varepsilon\sigma}$ and current $j_{\varepsilon\sigma}$:
\begin{align}
\frac{\p}{\p t}\rho_{\varepsilon\sigma} + \p_x j_{\varepsilon\sigma} = 0.
\end{align}
We can then read off:
\begin{align}
j_{\varepsilon\sigma} = \int^{\infty}_{-\infty}\frac{dp_x}{2\pi}p_x\frac{1}{e^{\sigma p_x/T_{\sigma}} + 1} - \bar j_{\varepsilon\sigma}.
\end{align}
The ground-state value $\bar j_{\varepsilon\sigma}$ can again be fixed by the normal ordering condition $j_{\varepsilon\sigma}(T_{\sigma} = 0) = 0$. We can therefore evaluate the energy current as follows:
\begin{align}
j_{\varepsilon\sigma} = \int^{\infty}_{-\infty}\frac{dp_x}{2\pi}p_x\left( \frac{1}{e^{\sigma p_x /T_{\sigma}} + 1} -\theta(-\sigma p_x)\right) = \frac{\sigma \pi T^2_{\sigma}}{12},
\end{align}
where $\theta$ is the standard step function. This computation implies the total anomalous energy current:
\begin{align}
j_{\varepsilon} = \sum_{\sigma}j_{\varepsilon\sigma} = \frac{\pi}{12}(T^2_+ - T^2_-).
\end{align}
To summarize, in this section, we established the kinetic approach for chiral anomaly in (1+1) dimensions using two methods: the anomalous commutation relation of charge densities and the gradient expansion of semiclassical microscopic Green's functions. We provided straightforward derivations of the anomalous number and energy currents as applications of our approach. In these derivations, we emphasized that the pivotal factor is the asymptotic behavior of the distribution function $n_{\sigma p}$ at $p_x\to\pm\infty$, rather than Berry phase as in (3+1) dimensions-- since Berry curvature cannot be defined in this scenario. 

Following this line of thought, it is worth commenting on whether the (1+1) dimensional analogue of the Berry phase, Zak's phase \cite{PhysRevLett.62.2747}, plays a role, as it is also relevant to charge polarization in solids \cite{RevModPhys.82.1959}. We have a couple of physical arguments suggesting that it does not. First, the definition of Zak's phase utilizes the periodicity of momentum space, meaning it is defined on a Brillouin zone. The arguments presented in this paper do not assume this property for the underlying spacetime. Moreover, within our scope, quantities appearing in the semiclassical Boltzmann equations are local and gauge invariant, while Zak's phase and its integrand do not satisfy these requirements.
}
\section{Universal chiral anomaly for odd dispersion}\label{dispersion_section}
Let us now shift focus and consider the quantum mechanical implications using the previous semi-classical argument. Although the Boltzmann equation was derived using a linear dispersion relation, the dependence on spectrum enters only indirectly through the asymptotic behavior of the particle number distribution function $n_{\sigma p}(x)$. By repeating the arguments from the previous sections and keeping the temporal derivative linear, a chirality-odd strictly monotonic dispersion relation should suffice for a model to exhibit chiral anomaly, regardless of the spacetime symmetry. The universality of chiral anomaly manifests in the sense that it is not sensitive to the precise form of the quasiparticle dispersion relation.
To prove this statement, we consider a power-law dispersion relation with an odd power:
\begin{align}
\label{oddModel}\mathscr L_{\ell} = \psi^{\dag}(iD_0 - \kappa\sigma^z (-iD_1)^{\ell})\psi, 
\end{align}
where $\ell$ is an odd integer and $\kappa$ is a dimensionful parameter to restore the unit of the term. Lagrangian~\eqref{oddModel} is invariant under the U(1) transformation $\psi \to e^{i\alpha}\psi$ and the chiral transformation $\psi \to e^{i\ga_5\alpha}\psi = e^{i\sigma_z\alpha}\psi$. To evaluate the chiral anomaly, let us deploy Fujikawa's recipe  \cite{PhysRevLett.42.1195, PhysRevD.21.2848, Fujikawa2004}, considering the partition functional
\begin{align}
\mathcal Z = \int \mathscr D\psi \mathscr D\bar{\psi}\, \exp\left(i\int d^2x\, \mathscr L_{\ell}[\psi, \bar{\psi}]\right).
\end{align}
Under a local chiral transformation, $\psi \to (1+i\alpha(x)\ga_5)\psi$, the chiral current emerges in the Lagrangian $\mathscr L_{\ell} + \alpha(x)\p_{\mu}j_5^{\mu}$, whereas the path integral measure is scaled by the Jacobian 
\begin{align}
\mathscr D\psi\mathscr D\bar{\psi} \to \mathscr J^{-2}\mathscr D\psi\mathscr D\bar{\psi},
\end{align}
where the logarithm of $\mathscr J$ involves evaluating the trace of the chiral operator over the Hilbert space. Using the standard Gaussian regularization, we have:
\begin{align}
\ln \mathscr J = i\int d^2x\, \alpha(x)\lim_{M\to\infty}\lan x| \mathrm{Tr}[\ga_5 e^{-\hat H^2/M^2} ]| x\ran,
\end{align}
where the {\it energy} operator $\hat H^2$ is given by:
\begin{align}
& -\hat H^2 = [\ga^0 iD_0 - \kappa\ga^1(-iD_1)^{\ell}]^2\notag\\
= & -D_0^2 -\kappa^2(-iD_1)^{2\ell} - \kappa\ga_5 iF_{01}\ell(-iD_1)^{\ell-1}.
\end{align}
Expanding the Fujikawa Jacobian to the first non-vanishing order and evaluating the trace, we see the amount of anomaly is independent of $\kappa$ and $\ell$:
\begin{align}
& \mathscr J =  -2i\kappa F_{01}\ell\lan x| e^{-[D_0^2 -\kappa^2(-iD_1)^{2\ell}]/M^2}\frac{(-iD_1)^{\ell-1}}{M^2}|x\ran\notag\\
= & 2 F_{01}\int \frac{d^2k}{(2\pi)^2} e^{-k_0^2 -\kappa^2k_1^{2\ell}}\ell \kappa k_1^{\ell-1} = \frac{1}{2\pi}F_{01}.
\end{align}
Note that a Wick rotation is performed on the temporal momentum so the integral is properly Euclidean. When evaluating the trace, it becomes evident that no anomaly can persist for even $\ell$ since the momentum integral simply vanishes. Similarly, this derivation straightforwardly generalizes to more generic dispersion relations as long as it is a strictly increasing function of the covariant derivative $\omega \sim \epsilon(-i\p_x + A_1)$. By inverse function theorem, the integrand and the integral domain can be bijectively mapped to the case of linear dispersion. Consequently, we have shown that the chiral anomaly in (1+1) dimensions is not sensitive to the exact shape of the dispersion. It is sufficient to have a strictly monotonic energy dispersion that is odd under chiral transformation in the fermion Lagrangian density in order to produce the chiral anomaly \eqref{anomaly_0}. The leading contribution has the universal magnitude $E/\pi$, and an extra minus sign may emerge if the dispersion is not strictly increasing in momentum. Note that in the case where $\epsilon$ is strictly decreasing, the notion of the left-moving and right-moving fermions are swapped. 
\section{Conclusion}
To recap, we demonstrated that in (1+1) dimensions, a model does not require an anomalous new term in its kinetic equation to exhibit chiral anomaly, unlike in (3+1) dimensions where the Berry phase associated with each Weyl fermion is essential. Instead, the asymptotic behaviors of the dispersion relations of the chiral fermions fully determine the value of momentum integral. Exploiting this observation, we provided an extremely simple derivation of the classic exercise of chiral anomaly in (1+1) dimensions, circumventing the subtle infinities of the ground state.  {We demonstrated that the anomalous number and energy currents in (1+1) dimensions can be easily derived using the kinetic equation.} Furthermore, we showed that a broader class of models could exhibit the same chiral anomaly.

It is physically intriguing to investigate the generic principles and forms of anomalies for generalizations of (1+1)-dimensional fermions with higher-order derivative couplings, including those considered in Refs. \cite{Barcelos-Neto1991, PhysRevD.47.3443}, where kinetic terms with higher-order derivatives are considered while preserving Lorentz symmetry. Mathematically, these models may be related to or motivate other index theorems beyond the scope of the Atiyah-Singer index theorem \cite{10.2307/1970715, Atiyah1973}.

Finally, it is worth noting that we deliberately kept the electromagnetic field classical and turned off short-range interactions for fermions mainly to streamline the derivation with minimal components. Exploring the robustness of our main results against interactions could be an intriguing follow-up exercise. Naively, some key arguments fail when turning on a four-fermion term, as the theory resembles a Luttinger liquid, casting doubt on putting the Green's function on-shell \cite{kamenev_2011}. Nevertheless, drawing insights from bosonization \cite{MANTON1985220}, a contact interaction should primarily renormalize the velocity of the dual boson field rather than its mass, which governs the amount of chiral anomaly in that formulation. If chiral anomaly persists with the same value in this context, it would indicate an alternative pathway for its manifestation in the classical limit.
\begin{acknowledgments}
We thank Umang Mehta and Hart Goldman for discussions and Professor Yeong-Chuan Kao for comments on the foundational ideas underlying this manuscript and for bringing the work in Ref.\onlinecite{PhysRevD.97.016018} to our attention. CHW is supported, in part, by ARO (W911NF-18-1-0020, W911NF-18-1-0212), AFOSR MURI (FA9550-19-1-0399), AFRL (FA8649-21-P-0781), and NSTC (111-2112-M-002-049-MY3, 111-2119-M-007-009-, and 112-2119-M-007-008-).
\end{acknowledgments}

\bibliography{citation}

\end{document}